\documentclass[aps,prl,twocolumn,noshowpacs,nofootinbib]{revtex4}

\usepackage{amsmath}
\usepackage{amsfonts}
\usepackage{amssymb}
\usepackage{amsthm}
\usepackage{mathrsfs}
\usepackage{dsfont}
\usepackage{esint}
\usepackage{graphicx}
\usepackage{physics}

\DeclareMathAlphabet{\mathpzc}{OT1}{pzc}{m}{it}

	\newcommand{\AsymEq}{\sim}

	\newcommand{\mr}[1]{\mathrm{#1}}			

	\newcommand{\sbr}[1]{( #1 )}
	\newcommand{\sbrr}[1]{[ #1 ]}
	
	\newcommand{\sof}[1]{\!\sbr{#1}}
	\newcommand{\soff}[1]{\!\sbrr{#1}}

	\newcommand{\Sum}[2]{\sum\limits_{#1}^{#2}}
	\newcommand{\Prod}[2]{\prod\limits_{#1}^{#2}}

	\newcommand{\sSum}[2]{\sum_{#1}^{#2}}
	\newcommand{\sProd}[2]{\prod_{#1}^{#2}}

	\newcommand{\EA}[1]{\xpc{#1}}

	\newcommand{\xpc}[1]{\left\langle #1 \right\rangle}

	\newcommand{\sVar}[1]{\mr{Var}\soff{#1}}

	\newcommand{\sLandau}[1]{\mathpzc{O}\sof{#1}}

		\newcommand{\Id}{\mathds{1}}

		\newcommand{\sSpan}[1]{\mathrm{Span}\soff{#1}}

\usepackage{xcolor}

\newcommand{\Ham}{\hat{H}}
\newcommand{\TEO}{\hat{U}}
\newcommand{\PsiIn}{\psi_\text{in}}
\newcommand{\TDP}{P_\text{det}}
\newcommand{\Detect}{\hat{D}}
\newcommand{\Hilbert}{\mathcal{H}}
\newcommand{\PsiDet}{d}

\begin{document}

\title{Uncertainty and symmetry bounds for the quantum total detection probability}
\author{Felix Thiel}
\author{Itay Mualem}
\author{David A. Kessler}
\author{Eli Barkai}
\affiliation{Department of Physics, Institute of Nanotechnology and Advanced Materials, Bar-Ilan University, Ramat-Gan
52900, Israel}

\begin{abstract}
  We investigate a generic discrete quantum system prepared in state $\ket{\PsiIn}$, under repeated detection attempts aimed to find the particle in state $\ket{d}$, for example a quantum walker on a finite graph searching for a node.
  For the corresponding classical random walk, the total detection probability $\TDP$  is unity. 
  Due to destructive interference, one may find initial states $\ket{\PsiIn}$ with $\TDP<1$. 
  We first  obtain an uncertainty relation which yields insight on this  deviation from classical behavior, showing the  relation  between $\TDP$  and  energy fluctuations: 
  $ \Delta P \,\sVar{\Ham}_d \ge \abs*{\mel*{\PsiDet}{ \comm*{\Ham}{\Detect} }{\PsiIn} }^2$ where $\Delta P = \TDP - \abs*{ \ip*{\PsiIn}{\PsiDet}}^2$, and $\Detect = \dyad{\PsiDet}$ is the measurement projector. 
  Secondly, exploiting symmetry we show that $\TDP\le 1/\nu$ where the integer $\nu$ is the number of states equivalent to the initial state. 
  These bounds are compared with the exact solution for small systems, obtained from an analysis of the dark and bright subspaces, showing the usefulness of the approach.  
  The upper bounds works well even in large systems, and we show how to tighten the lower bound  in this case.
\end{abstract}

\maketitle

The dynamics of quantum systems that evolve unitarily but are subject to repeated monitoring 
using projective measurements have gained recent attention partly driven by increasing interest in quantum information \cite{Krovi2006-0, Krovi2006-1, Krovi2007-0, Varbanov2008-0, Gruenbaum2013-0, Bourgain2014-0, Dhar2015-0, Dhar2015-1, Lahiri2019-0,  Sinkovicz2015-0, Sinkovicz2016-0, Friedman2017-1, Thiel2018-0, Thiel2018-1, Kay2010-0, Novo2015-0, Chakraborty2016-0, Mukherjee2018-0, Rose2018-0}.
The investigation of a single particle on a finite graph, i.e. a quantum walker \cite{%
  Muelken2006-0, Perets2008-0, Karski2009-0, Zaehringer2010-0, Muelken2011-0, Venegas-Andraca2012-0,%
  Preiss2015-0, Xue2015-0%
}, prepared and detected in the states $\ket{\PsiIn}$ and $\ket{d}$, respectively, was promoted as 
a basic model in the context of quantum search \cite{Grover1997-0, Farhi1998-0, Childs2002-0, %
  Aaronson2003-0, Bach2004-0, Childs2004-0, Kempe2005-0, Muelken2006-0, Magniez2011-0, %
  Krovi2006-0, Krovi2006-1, Varbanov2008-0, Li2017-0%
}.
Both states may (or may not) be localized in the graph node basis.
The key quantifier of this process of unitary evolution mingled with wave function collapse is the total detection probability $\TDP$.
This is the fraction of particles in the statistical ensemble which are eventually detected.
Previous work \cite{Plenio1998-0, Facchi2003-0, Facchi2008-0, Krovi2006-0, Caruso2009-0} showed how 
non-detectable initial states ({\em dark} states) may render the quantum search impossible, such that $\TDP(\PsiIn) =0$.
Likewise, a general initial condition may yield a total detection probability $\TDP$ less than unity, 
in stark contrast to a classical random search on the same structure for which $\TDP$ is always one \cite{Redner2007-0, Boettcher2015-0}.
However, while $\TDP$ is known explicitly for some specific examples \cite{Krovi2006-0, Friedman2017-1}, 
general principles leading to its estimation are still missing.
In this Letter we provide three insights on $\TDP$: an uncertainty principle, symmetry arguments and an exact solution.

\begin{figure}
  \includegraphics[width=0.9\columnwidth]{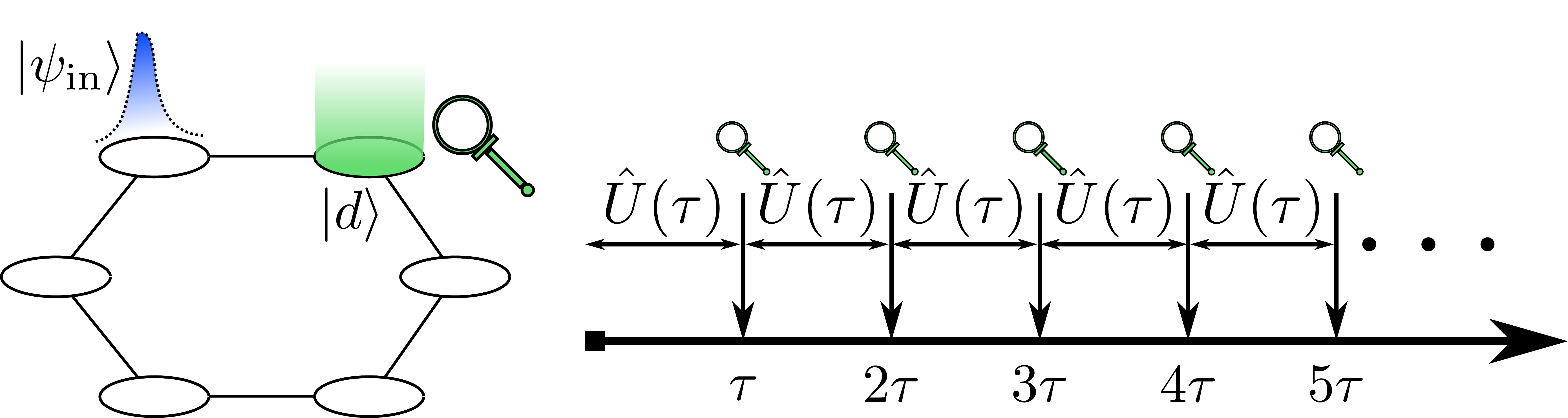}
  \caption{
    A quantum walker resides on the nodes of a graph and moves unitarily along its edges, here a ring (left).
    Every $\tau$ time units a detector tests whether the particle is at node $\ket{d}$ collapsing the wave function (right).
    The first successful detection attempt (click) defines an arrival time and stops the protocol.
    $\TDP$ is the probability that the detector clicks at all.
    Here, the initial state is localized.
    \label{fig:Model}
  }
\end{figure}

The exact solution relies on decomposing the Hilbert space into mutually orthogonal dark and bright sub-spaces.
These are examples of Zeno subspaces \cite{Facchi2008-0, Caruso2009-0, Mueller2017-0}, a dynamical separation
of the total Hilbert space, which usually appears in the presence of singular coupling or in {\em rapidly} measured systems.
Here, they are relevant for arbitrary detection frequencies possibly far away from
the regime of the quantum Zeno effect \cite{Misra1977-0, Itano1990-0}.
This formal solution requires a full diagonalization of the Hamiltonian, involving considerable effort.
Therefore, we also present bounds on $\TDP$, that give physical insight into the problem.
The lower bound is an uncertainty relation and the upper bound exploits symmetry. 

  Heisenberg's uncertainty relation is probably the most profound 
signature of
 quantum reality's deviations 
 from classical Newtonian mechanics \cite{Busch2007-0}. Here we  show  something
very different: how $\TDP$ of a  quantum walk deviates from
the corresponding probability  of detecting a classical random
walk, which is unity.  
Our uncertainty relation connects this deviation with energy fluctuations and the commutator of $\Ham$
and the measurement projector.
It follows from the collapse postulate.

 Symmetry and  degeneracy 
play an important part in the physics of dark states
and are  a crucial mechanism leading to $\TDP < 1$ \cite{Krovi2007-0}. 
Consider an initial state which is a superposition of two energy eigenstates,
$\ket{\PsiIn} = N ( \bra{d} E \rangle \ket{E'} - \bra{d} E' \rangle \ket{E})$. 
When $\ket{E}$ and $\ket{E'}$ belong to the same energy level, i.e. $\Ham\ket{E} = E \ket{E}$ and $\Ham\ket{E'} = E \ket{E'}$,
the time evolution of $\ket{\PsiIn}$ is a simple phase factor $e^{-itE}$ (here $\hbar=1$).
It follows that $\mel*{d}{e^{-it\Ham}}{\PsiIn} = 0$ forever. 
Hence, this is a dark state. 
Importantly, degeneracy is a signature of $\Ham$'s  symmetry,
so dark states are deeply connected to the symmetry of the problem.
Below we exploit this to find a simple bound on
$\TDP$.

\paragraph*{Model}  
We consider quantum systems with discrete states, e.g. quantum walks on finite graphs.
The particle is initially in state $\ket{\PsiIn}$.
We use projective stroboscopic measurements at times $\tau, 2 \tau, ... $ in an attempt
to detect the particle in state $\ket{d}$; see Fig.~\ref{fig:Model} and Refs.~\cite{Gruenbaum2013-0,Friedman2017-1,Thiel2018-0}. 
The detection could be performed  on a node
of the graph, though any state $\ket{d}$ is acceptable.
Between the measurement attempts the evolution is unitary, described with $\hat{U}(\tau)= \exponential( - i\tau \Ham)$.
The string of measurements yields a sequence no, no, $\cdots$  and in the $n$-th attempt a yes.
The time $n\tau$ marks the first detected arrival time in state $\ket{d}$.
In some measurement sequences, the particle is not detected at all ($n=\infty$).
Each  measurement satisfies the collapse postulate \cite{Cohen-Tannoudji2009-0}:
if the wave function is $\ket{\psi}$ right before measurement, the amplitude of detection is $\ip{d}{\psi}$.
Successful detection terminates the experiment.
Unsuccessful detection zeroes the amplitude $\ip{d}{\psi}$,
the wave function is renormalized, and the unitary evolution continues until the next measurement. 
Mathematically, the measurement is  described
by the projector $\Detect = \dyad{d}$ (see Eq.~\eqref{eqPHIN} below).  
Repeating this protocol many times, 
$\TDP$  is the fraction  of runs in which the detector clicked yes at all.
If $\TDP < 1$, the mean $\EA{n}$ diverges and the corresponding search problem is ill-posed.
Furthermore, the time-of-arrival problem \cite{Allcock1969-0, Allcock1969-1, Allcock1969-2, Echanobe2008-0, Ruschhaupt2009-0, Sombillo2016-0} can also be tackled using stroboscopic measurements.
General quantum walks have been experimentally realized photonically \cite{Perets2008-0, Schreiber2012-0, Xue2015-0}, with trapped ions \cite{Zaehringer2010-0} and in optical lattices \cite{Karski2009-0, Sherson2010-0}.
Ref.~\cite{Nitsche2018-0} reported the measurement of $\TDP$ in a photonic walk.

\paragraph*{Uncertainty relation}
Ref.~\cite{Krovi2006-0} showed how the detector's action separates the Hilbert space  
of a finite system into a ``bright'' and a ``dark'' subspace:
${\cal H} = {\cal H}_B \oplus {\cal H}_D$. 
Any initial condition  within the bright/dark subspace is detected
with probability one/zero respectively.
We present a
proof of this fundamental result 
in Ref.  \cite{Thiel2019-0}.
For the dark, as for the bright space, we can find a basis 
in terms of the eigenstates of $\Ham$, denoted
$\{ \ket{ E_j ^{B} }\}$  and $\{\ket{ E_j ^{D}}\}$ respectively. 
The subspaces are thus orthogonal and invariant under $\Ham$ and $\Detect$.

\begin{figure*}
\centering
\includegraphics[width=0.99\textwidth]{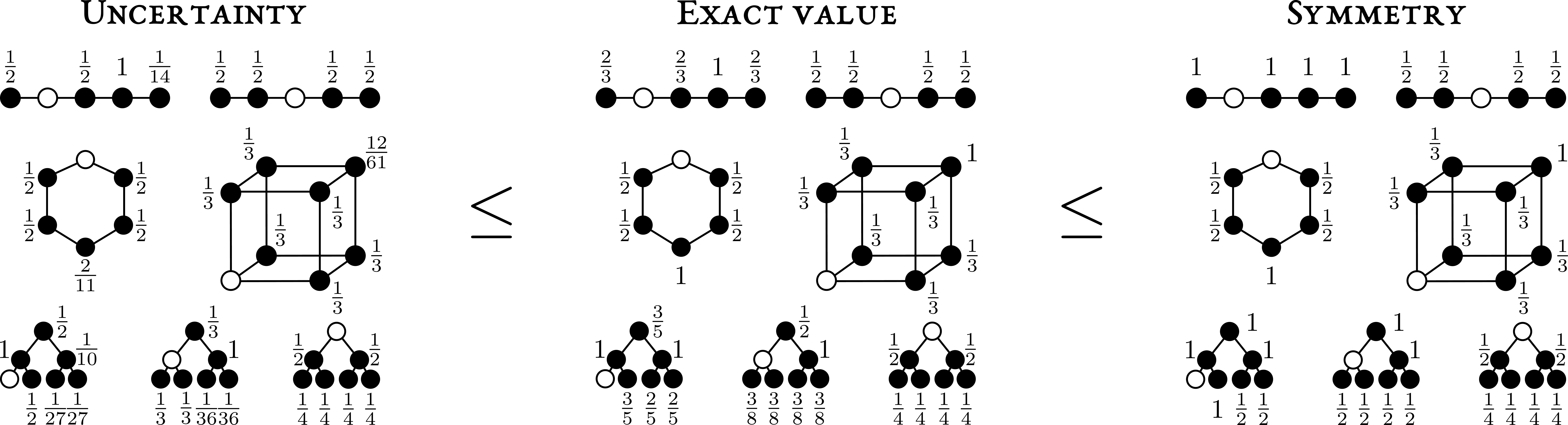}
\caption{
  Uncertainty bound, exact value and symmetry bound for $\TDP$.
  Each graph represents a quantum walk, the open circles denote 
  the position of the detector, the filled circles are possible localized initial states, the numbers are the bounds or values for $\TDP$.
  Left: Uncertainty bound, Eqs.~(\ref{eqs1}, \ref{eq:PathCount}).
  Middle: exact solution, Eq.~\eqref{eq03}.
  Right: Symmetry bound, Eq.~\eqref{eqUpper}.
  The Hamiltonians are equal to the graph's adjacency matrices.
  For the lower bound we took $s$ equal to the distance between initial and detection node.
  In many cases lower and upper bound coincide and $\TDP$ is determined without tedious calculations.
  For starred graphs the shell state method gives exact results.
}
\label{fig:Comp}
\end{figure*}

 Ref. \cite{Gruenbaum2013-0} (and Eq.~\eqref{eq03} below) showed that the particular state $\ket{\PsiIn}=\ket{d}$ is bright.
Since $\ket{d}$ is bright, it is orthogonal to every dark state,
i.e. $\bra{E_j ^{D}} d \rangle=0$. 
Therefore, $\Ham^s \ket{d}$ is also bright, because $\bra{E_j ^{D}} \Ham^s \ket{d}= (E_j) ^{s} \bra{E_j ^{D}} d \rangle=0$, where $s$ is any positive integer. 
$\TDP$ is the overlap of $\ket{\PsiIn}$ with $\mathcal{H}_B$, so given any orthonormal basis 
 $\{ \ket{\beta_l}\}$ of $\mathcal{H}_B$ we have:
\begin{equation}
\TDP = \sum_{l} |\bra{\beta_l} \PsiIn \rangle|^2  .
\label{eqSS}
\end{equation} 
To obtain a useful bound,
we create a pair of orthonormal bright states from $\ket{d}$ and $H^s\ket{d}$:
\begin{equation} 
 \ket{\beta_1} 
 = 
 \ket{d} 
 \qc 
 \ket{\beta_2} 
 = 
 N [ \Id - \dyad{\PsiDet} ] \Ham^s \ket{\PsiDet}
\label{eqb1b2}
\end{equation}
The normalization $\abs{N}^{-2} = \sVar{\Ham^s}_{\PsiDet} = \ev*{\Ham^{2s}}{\PsiDet} - [ \ev*{\Ham^s}{\PsiDet} ]^2$ is related to the energy fluctuations in the detected state.
As each term in the sum  Eq.~(\ref{eqSS}) is non-negative, a lower bound is reached by omitting some of the bright states:
\begin{equation}
 \TDP \ge | \bra{\beta_1} \PsiIn \rangle|^2  
+  | \bra{\beta_2} \PsiIn  \rangle|^2  . 
\label{eqchichi1}
\end{equation}
We now define the difference between the
probability of detection after repeated measurements from the initial
probability of detection:
\begin{equation}
 \Delta P = \TDP - | \ip{d}{\PsiIn}|^2. 
\end{equation}
Using Eqs. 
(\ref{eqb1b2},
\ref{eqchichi1}) we find
\begin{equation}
\Delta P \,
\sVar{\Ham^s}_d  
\ge  
|\bra{d} [ \Ham^s, \Detect ]  \ket{\PsiIn} |^2.
\label{eqs1}
\end{equation}
Fig.~\ref{fig:Comp} shows the uncertainty bound for several graphs.
Some remarks are in place.  
First, after the successful  detection, the system is in its final  state $\ket{\psi_\text{f}} =\ket{d}$\footnote{
  Here and all along $\ket{\psi}$ denotes the state of the system/particle 
  along the measurement process, e.g. in the initial of final state, while $\ket{d}$ is the detection state.
}.
This means that we may rewrite the uncertainty principle, say for $s=1$, as
\begin{equation}
  \Delta P \, \sVar{\Ham}_{\psi_\text{f}}  
  \ge 
  \abs*{ \mel*{\psi_\text{f}}{ \comm*{ \Ham}{ \Detect }}{\psi_\text{in}}}^2,
\label{eqnewRob}
\end{equation}
The fluctuations of energy are actually in the {\em final state} of the particle.
So, Eqs.~(\ref{eqs1}, \ref{eqnewRob}) are relations between the initial condition and the finally selected state.
Importantly, after the system is projected into its final state, and the detector turned off, 
the fluctuation of energy $\sVar{\Ham}_{\psi_\text{f}}$ is a constant of motion.
The energy measurement can be made at any time after the detection. 
Notably, Eqs.~(\ref{eqs1},~\ref{eqnewRob}) do not depend on $\tau$.

\paragraph*{Path-counting approach} We consider the standard
quantum walk with  $\Ham=\hat{A}$, where $\hat{A}$ is the adjacency matrix
 of some  graph.
Hence, there are no on-site energies, and all bonds
in the system are identical, namely $\Ham_{ii}=0$ and $\Ham_{ij}=1$ if
site $i$ and $j$ connected, zero otherwise.
We are interested in a particle starting on vertex
$\ket{\PsiIn}=\ket{r}$ and the detection on another
vertex $\ket{d}$. 
Notice that $\bra{d} \Ham^s \ket{r}= {\cal N}_{r \to d} (s)$ is the number of paths of length $s$ starting on $\ket{r}$ and ending at $\ket{d}$.
Then using Eq.~(\ref{eqs1}), we find:
\begin{equation}
\TDP(r ) \ge { |{\cal N}_{r \to d} (s)|^2
\over {\cal N}_{d \to d} (2 s) - \left[ {\cal N}_{d\to d} (s)\right]^2 }.
\label{eq:PathCount}
\end{equation}
We must choose $s$ here larger or equal to the distance $\xi$ between $\ket{r}$ and $\ket{d}$,
otherwise one gets the trivial $\TDP \ge 0$.

\paragraph*{Upper bound from symmetry} 
To complement our lower bound, we use a different approach.
 The detection
probability is by definition $\TDP= \sum_{n=1} ^\infty |\varphi_n|^2$
where $\varphi_n$ is the amplitude of first detection at the $n$-th attempt \cite{Friedman2017-1}.
This can be expressed as \cite{Dhar2015-1}:
\begin{equation}
\varphi_n = \bra{d} \hat{U}(\tau)[ ( \Id - \Detect ) \hat{U}(\tau) ]^{n-1} \ket{\PsiIn }. 
\label{eqPHIN}
\end{equation}
Reading this right-to-left, we see that $\varphi_n$
is given by the initial condition, followed by steps combining unitary evolution and attempted detection,
of which the final, $n$-th detection is successful. 
It  is crucial for our discussion  that $\varphi_n$
 is
linear with respect to $\ket{\PsiIn}$, so it obeys 
the superposition
principle.

 We are interested in the total detection probability starting
 from node $\ket{r}$ and detecting on  another
$\ket{d}$. In the system we have a set 
$\left\{ \ket{r_j} \right\}_{j=1} ^\nu$  
of $\nu$  states
which are
equivalent
to  $\ket{r}$
and $\ket{r_1} = \ket{r}$.
This means that each $\ket{r_j}$ gives the same amplitude on
$\ket{d}$ for all times, mathematically
$\bra{d} \hat{U}(t) \ket{r} = \bra{d} \hat{U}(t) \ket{r_j}$  
for $1\le j \le \nu$. 
Physically, it is often  easy to identify 
all the states $\ket{r_j}$ using symmetry arguments.
However, even if we miss some of them,
the bound derived below is useful though not optimal.  

 From  the equivalent states $\ket{r_j}$, we construct 
 a normalized  auxiliary uniform  state
\begin{equation}
\ket{{\rm AUS}} := \frac{\sum_{j=1} ^\nu  \ket{r_j} }{ \sqrt{\nu} }.
\end{equation}
Now, by definition of the detection amplitudes 
and the equivalence of  all 
$\left\{ \ket{r_j} \right\}_{j=1} ^\nu$,  
we find $\varphi_n (r_j)=
\varphi_n (r)$.
It follows from superposition, Eq.~(\ref{eqPHIN}), that
\begin{equation}
\varphi_n ({\rm AUS}) = \sqrt{\nu} \varphi_n (r ). 
\end{equation}
We now square both sides of this equation, sum over $n$, 
and use the obvious $\TDP({\rm AUS}) \le 1$ to find the sought after $\TDP(r)$:
\begin{equation}
\TDP(r ) = {\TDP ({\rm AUS} )\over \nu} \le {1 \over \nu}.
\label{eqUpper}
\end{equation}
Fig.~\ref{fig:Comp} shows the upper bound for several graphs.

If the system is disordered, then generically $\nu=1$ and the inequality saturates, as shown below.
Thus, symmetry yields a useful bound and disorder gives essentially classical behavior $\TDP= 1$, see also \cite{Caruso2009-0}. 
Ref.~\cite{Thiel2019-1} will show that $\nu = \dim \mathcal{S}_d\ket{\PsiIn}$ can be determined from the stabilizer ${\cal S}_d$, the group of all symmetry operations that commute with $\TEO(\tau)$ and $\Detect$.

\paragraph*{Ring}
Consider a ring with an even number $L$ of identical sites, with localized initial and detection states.
The detection site and its opposing site are unique, such that $\nu=1$. 
For all other sites, we have  one equivalent partner found by reflection
symmetry, hence $\nu=2$. 
In the supplementary material (SM), we derive a lower bound from Eq.~\eqref{eq:PathCount} with $s=\xi<L/2$:
\begin{equation}
  \frac{1}{\binom{2\xi}{\xi} - [\binom{\xi}{\xi/2}]^2}
  \le \TDP(d\pm\xi) \le \frac{1}{2}
  ,
\label{eq:}
\end{equation}
where the second binomial must be omitted for odd $\xi$.
For nearest neighbors $\xi=1$, we find the exact result $\TDP(d\pm1) = 1/2$ from sandwiching.
Consider now the detection of the ring's ground state $\ket{d} = \sSum{r=1}{L} \ket{r}/\sqrt{L}$.
Since $\ket{d}$ is a uniform state over the whole ring, each localized initial condition is physically equivalent.
The upper bound gives $\TDP(r) \le 1/L$, which is also equal to the exact result.

\paragraph*{The exact solution is} 
\begin{equation}
\TDP =
\sideset{}{'}\sum_{l}
{ \left| \sum_{m=1} ^{g_l} \bra{d} E_{l,m} \rangle\bra{E_{l,m}}
\PsiIn\rangle \right|^2 \over
\sum_{m=1} ^{g_l} \left|\bra{d} E_{l,m}\rangle\right|^2}.
\label{eq03}
\end{equation}
Here  the eigenstates $\ket{E_{l,m}}$ and the energies $E_l$ 
are defined as usual with
  $\Ham\ket{E_{l,m}} = E_l \ket{E_{l,m}}$ where
$l,m$ are quantum numbers, and  $m=1,...,g_l$  so
$g_l$ is the degeneracy of energy level $E_l$.
The sum runs over all $l$ for which the denominator does not vanish.
Let us briefly outline the derivation of this formula
and then discuss its consequences.

\paragraph*{Sketch of proof}
Eq. (\ref{eq03}) follows directly from the decomposition of the Hilbert space into dark and bright components. 
Technically, we use the energy basis and consider an energy sector $\{ \ket{E_{l,m}} \}_{m=1} ^{g_l}$.
This sector yields either one bright state (and $g_l-1$ dark states)
 or none at all (and $g_l$ dark states). 
If $\bra{E_{l,m} } d \rangle=0$ for all $1 \le m \le g_l$
then clearly all the $g_l$  states are dark and the sector has no bright state.
Otherwise, there is only one bright state, namely $\ket*{E_l^B} = N_l ^{B} \sum_{m=1} ^{g_l} \ket{E_{l,m}}  \bra{E_{l,m}} d \rangle$ with appropriate normalization.
We need to demonstrate: (i)  that
indeed $\ket*{E_l^B}$ is bright, 
and (ii)  that the remaining states are dark. 
The latter is easily shown.
Consider for example $g_l=2$. We have  
$\ket*{E_l^{B} } = N_l ^{B} \left( \bra{E_{l,1}} d \rangle\ket{E_{l,1}}  + 
 \bra{E_{l,2}} d \rangle\ket{E_{l,2}} \right)$ . It is easy
to see that 
$\ket*{E_l^{D} } = N_l ^{D} \left( \bra{E_{l,2}} d \rangle\ket{E_{l,1}}  - 
 \bra{E_{l,1}} d \rangle\ket{E_{l,2}} \right)$  is dark as
 $\bra{d}E_l ^{D} \rangle =0$ and $\ip*{E_l^B}{E_l^D} = 0$.
Similar arguments  hold for $g_l > 2$ \cite{Thiel2019-0}. 
Showing that $\TDP(E_l^B) = 1$ is involved.
For that aim, we analyzed in Ref. \cite{Thiel2019-0} 
 the eigenvalues of the operator $(\Id-\Detect) \hat{U}(\tau)$ which
determine the evolution of the measurement process.
These eigenvalues lie inside the unit disk.
This fact is used to show that $\ket*{E^B_l}$ is detected with probability one. Once we have all the
bright states, we use Eq.  
(\ref{eqSS})
to obtain Eq. 
(\ref{eq03}).

\paragraph*{Features of Eq.~\eqref{eq03}}
The exact formula exhibits some remarkable
properties. The first is that
the  detection probability is $\tau$-independent.
The only exception, not considered here in depth, 
 is when $|E_l - E_{l'} | \tau = 0\  \mbox{mod} \ 2 \pi$,
for some pairs of energy levels. 
They are a unique feature of the stroboscopic detection protocol.
These special $\tau$s are isolated, but still of interest since the statistics 
 exhibit gigantic
fluctuations and discontinuous behavior in their vicinity \cite{Gruenbaum2013-0,Yin2019-0}, 
related to partial revivals of the 
state function. More importantly, Eq.~\eqref{eq03}'s $\tau$-independence ensures its general validity,
even if one tampers with the detection
protocol; for example by sampling with a Poisson process. 
The reason is that any
initial state $\ket{\PsiIn}$ starting in the dark space
has zero overlap with the detected state for $t\ge 0$.
No measurement protocol can detect this state. 
Secondly, using Eq.~(\ref{eq03}), it is now easy to
see that a finite
 disordered system exhibits a classical behavior, as mentioned.
Namely,  if the system has no degeneracy and all the eigenstates
have a non-vanishing overlap with the detector, we find
$\TDP=\bra{\PsiIn} \PsiIn\rangle =1$.
Hence, disorder increases the probability of detection,
and a disordered system behaves like a random walk.
Finally,  for the return problem $\ket{\PsiIn} =\ket{d}$, and for $\ket{\PsiIn} = N_s \Ham^s\ket{d}$,
we get $\TDP=1$ from Eq.~(\ref{eq03}).
Hence, as claimed earlier, the states $\ket{d}$ and $\Ham^s \ket{d}$ are bright. 

Fig.~\ref{fig:Comp} compares our main results, the uncertainty relation (\ref{eqs1}), and the symmetry bound (\ref{eqUpper}), 
with the exact result (\ref{eq03}). 
In some cases both bounds coincide and thus determine the exact results.
Only elementary calculations -- in contrast to full diagonalization of $\Ham$ -- 
are necessary to obtain the bounds.
Contemporary quantum walk experiments achieve up to fifty steps before they decohere \cite{Nitsche2018-0}, hence our focus on small systems.

\paragraph{Large Systems} 
  Are the bounds found here useful for large systems as well? 
  We consider this question in the context of the $B$-dimensional hypercube.
  Each node is represented by a string of $B$ bits, e.g., $\ket{ 01011\cdots 0}$, and each transition corresponds to flipping one bit. 
  We detect on node $\ket{d}$ and start at any node $\ket{r_\xi}$ with $\xi$ bits different from $\ket{d}$, i.e., $\xi$ is the Hamming distance between the nodes. 
  Remarkably, the upper bound works perfectly, coinciding with the exact result $\TDP(r_\xi) = 1/\nu = 1/\binom{B}{\xi}$, and so the symmetry-related upper bound is seen to work well for large systems as well as small.
  What about the lower bound? 
  Eq.~\eqref{eq:PathCount} yields (see SM):
  \begin{equation}
    \frac{\xi!^2 B^{-\xi}}{(2\xi-1)!!-[(\xi-1)!!]^2} \le \TDP(r_\xi) 
    = 
    \frac{1}{\nu}
    =
    \frac{1}{{B \choose \xi}}
    .
  \end{equation}
  The second term in the denominator has to be omitted when $\xi$ is odd.
  When $B$ and $\xi$ are both large and comparable, the lower bound can be seen to decay exponentially in $B$.
  Thus it clearly needs improvement for larger systems when $\xi$ is not small.

  For $\xi$ near $B$, a very tight bound can be obtained  upon the realization that $\ket{r_B}$, the one node with maximal distance to $\ket{d}$, is bright. 
  Then one chooses the bright states $\ket{r_B}$ and $\Ham^{B-\xi}\ket{r_B}$ in Eq.~\eqref{eqb1b2} and obtains a lower bound that performs well, where the original one fails (see SM for details).

  This last approach succeeds because one of the two states used to generate the bound has a large overlap with the initial state. 
  A more general approach to achieve this end is based on the natural shell structure of the graph. 
  When the bright state  $\ket{d}$ is localized, $\Ham\ket{d}$ is supported only on nearest neighbors of $\ket{d}$. 
  We call those nodes the ``first shell''. 
  Similarly $\Ham^2 \ket{d}$ is a bright state supported on the next-nearest neighbors, the second shell, as well as $\ket{d}$ itself. 
  Since the $s$-th shell is only connected to the $s\pm1$-th shells, we can construct a useful bright state $\ket*{\tilde{\beta}_\xi }$ with maximal overlap with the initial state by the following strategy: 
  We start with the zeroth and first shell states $\ket*{\tilde{\beta}_0} :=\ket{d}$ and $\ket*{\tilde{\beta}_1} :=\Ham\ket{d}$. 
  Each subsequent state $\ket*{\tilde{\beta}_s}$ is obtained from orthogonalizing $\Ham\ket*{\tilde{\beta}_s}$ to   $\ket*{\tilde{\beta}_{s-1}}$. 
  The procedure is terminated  when $s=\xi$, and yields a state supported only in the $\xi$-th shell.
  A lower bound is obtained from $\TDP(r_\xi) \ge \abs*{\ip*{\tilde{\beta}_\xi}{\PsiIn}}^2$.
  For our nearest-neighbor hopping $\Ham$ on the hypercube, $\ket*{\tilde{\beta}_\xi} = \sum_{r_\xi} \ket{ r_\xi}/\nu$, where $\nu = {B \choose \xi}$, which is nothing but the relevant AUS that appeared in our symmetry-derived upper bound.
  Hence, 
  \begin{equation}
    \frac{1}{\nu}  \le \TDP(r_\xi) \le \frac{1}{\nu},
  \end{equation}
  and the lower and upper bounds coincide, yielding the exact result.

  In many situations (e.g. those starred in Fig.~\ref{fig:Comp}), this procedure gives the exact, simply computed, result.
  It is exact for systems where the sequence $\{\ket*{\tilde{\beta}_s}\}$ turns out to be fully orthogonalized. 
  We will elaborate on this method and when it is exact in an upcoming publication \cite{Thiel2019-2}. 
  Even when it is not exact, it gives a good lower bound that only involves $2\xi-1$ orthogonalization operations. 
  This is a huge advantage compared to the minimally necessary $\xi(\xi+1)/2$ operations in a full Gram-Schmidt procedure \cite{Novo2015-0}.

  To conclude, we have used symmetry and an uncertainty principle to find upper and lower bounds on the detection probability. 
  These bounds show a symmetry-induced restriction to well-posed quantum search which requires $\TDP = 1$.
  Due to the strong connection between stroboscopic detection and non-Hermitean models, our results are also relevant for continuous-time models \cite{Caruso2009-0, Ruschhaupt2009-0, Novo2015-0, Dhar2015-0}.
  Surprisingly, $\TDP$ is almost independent of $\tau$.
  The uncertainty principle is generally valid, but its usefulness is limited when there are many bright states.
  Starting from the hypercube example, we presented the generally applicable shell-state method that greatly improves the uncertainty relation.
  It gives a bright state which is heavily overlapping with the corresponding AUS, resulting in a tight sandwich for $\TDP$ even for large systems.
  While the exact result for $\TDP$ can always be used, it demands the full diagonalization of the problem, something that should be avoided if possible, and is harder to interpret physically.

\begin{acknowledgments}
  The support of Israel Science Foundation's grant 1898/17 is acknowledged.
  FT is endorsed by Deutsche Forschungsgemeinschaft (Germany) under grant number TH 2192/1-1.
\end{acknowledgments}

\appendix
\section{Counting paths in the ring}
  Here we derive a lower bound for the ring with $L$ sites from Eq.~(7) of the main text.
  Let the detection site be $d$ and initial state localized in $r = d+\xi$.
  We assume that $r$ is not the site opposite of $d$, i.e. $\xi < L/2$.
  We have to find the paths from the initial site to the detection site $\mathcal{N}_{r\to d}(s)$ and the number of returning paths $\mathcal{N}_{d\to d}(s)$.
  We consider the simplest non-vanishing lower bound with $s=\xi$.

  Returning in $s < L$ steps necessitates $s/2$ steps in each direction in arbitrary order.
  Hence ${\cal N}_{d\to d}(2\xi) = \binom{2\xi}{\xi}$ and also ${\cal N}_{d\to d}(\xi) = \binom{\xi}{\xi/2}$, provided $\xi$ is even.
  On the other hand there is only one path from $d+\xi$ to $d$ in $\xi<L/2$ steps, hence ${\cal N}_{r\to d}(\xi) =1$.
  Plugging these quantities into Eq.~(7) of the main text yields the lower bound of Eq.~(12).

\section{Counting paths in the hypercube}
  To evaluate Eq.~(7) of the main text for the hypercube, we need to compute ${\cal N}_{r_\xi\to d}(s)$ and ${\cal N}_{d\to d}(s)$.
  This is done using the bit representation.
  Take $s=\xi$.
  Reaching $\ket{d}$ from $\ket{r_\xi}$ necessitates $\xi$ bit flips in arbitrary order, hence $\mathcal{N}_{r_\xi\to d}(\xi) = \xi!$.
  A return from $\ket{d}$ to $\ket{d}$ in $s$ steps is only possible, if $s$ is even.
  It involves flipping each bit an even number $s_j$ times, such that $\sSum{j=0}{B} s_j = s$.
  There are $s! / (s_1! \cdots s_B!)$ combinations that this happens.
  (This is the multinomial coefficient.)
  The requirement of even  $s_j$ can be expressed via $[1 + (-1)^{s_j}]/2$.
  Hence, we have:
  \begin{equation}
    {\cal N}_{d \to d}(s)
    =
    \Sum{s = \sSum{j=1}{B} s_j}{} 
    \frac{s!}{\sProd{j=1}{B} s_j!}
    \Prod{j=1}{B} \frac{1 + (-1)^{s_j}}{2}
  \end{equation}
  When the parentheses are factored out, we use the generalized binomial theorem which states that $(x_1 + x_2 + \cdots + x_B)^s = \sSum{\{s_j\}}{} x_1^{s_1} \cdots x_B^{s_B}  s!/ (s_1! \cdots s_B!)$.
  Then we find that there are $\binom{B}{l}$ terms with $l$ factors $(-1)$.
  This gives 
  \begin{equation}
    {\cal N}_{d \to d}(s)
    =
    \frac{1}{2^B} \Sum{l=0}{B} \binom{B}{l} ( B - 2l )^s
    =
    2^s \EA{(\EA{R_B} - R_B)^s}
  \label{eq:}
  \end{equation}
  The remaining expression is proportional to the $s$-th central moment of a binomial random variable $R_B$ with $p=1/2$.
  When $B$ is large, the DeMoivre-Laplace limit theorem allows to replace $R_B \AsymEq (B/2) + Z \sqrt{B/4}$ with a normal random variable $Z$.
  The central moments of a Gaussian variable are known $\EA{Z^s} = (s-1)!!$ and they vanish when $s$ is odd (as does the exact expression).
  $(2k+1)!! = 1\cdot 3 \cdot 5 \cdots (2k+1)$ is the double factorial.
  Therefore, we find:
  \begin{equation}
    {\cal N}_{d\to d}(s)
    =
    \left\{ \begin{split}
      B^{s/2} (s-1)!!\qc & \text{ even } s \\
      0 \qc & \text{ odd } s \\
    \end{split} \right.
    .
  \label{eq:}
  \end{equation}
  Together with ${\cal N}_{r_\xi \to d}(\xi) = \xi!$ in Eq.~(7) of the main text, we obtain one part of Eq.~(14) of the main text:
  \begin{equation}
    \TDP(r_\xi) \ge \frac{\xi!^2 B^{-\xi}}{(2\xi-1)!! - [(\xi-1)!!]^2}
    .
  \label{eq:}
  \end{equation}
  Two limits should be discussed here.
  First, when $B$ is large but $\xi$ is finite, the lower bound decays like $B^{-\xi}$.
  This has to be compared with the dimension of the total Hilbert space $D = 2^B$.
  Our bound only decays as a power of $D$'s logarithm, that is moderately.
  This, however, breaks down when $\xi \sim B$ is comparable to $B$.
  Then, using $(2\xi-1)!! = (2\xi)! / (2^\xi \xi!)$ and a stirling approximation yields:
  \begin{equation}
    \TDP(r_\xi) \ge \frac{\xi!^2 B^{-\xi}}{(2\xi-1)!! - [(\xi-1)!!]^2}
    \AsymEq
    D^{- \frac{\xi}{B} \log_2 \frac{2eB}{\xi} }
    .
  \label{eq:}
  \end{equation}
  In this case the lower bound decays as a power of the (gigantic) system size.
  This motivated our search for more powerful lower bounds, using the shell method.

\section{Shell state method}
  Let us briefly review how to obtain $\TDP$ formally from the Gram-Schmidt procedure.
  The exact value of $\TDP(\PsiIn)$ is given by the overlap with the bright space $\Hilbert_B = \sSpan{\Ham^k\ket{\PsiDet} \, | \,  k=0,1,\hdots }$.
  Having found a basis $\{ \ket{\beta_k} \}_{k=0}^{w-1}$for $\Hilbert_B$, we can compute $\TDP(\PsiIn) = \sSum{k=0}{w-1} \abs*{\ip*{\beta_k}{\PsiIn}}^2$ as in the main text.
  $w = \dim \Hilbert_B$ is the maximum number of linearly independent states in the bright space.
  When the states $\ket{\beta_k}$ are determined from a Gram-Schmidt procedure applied to the sequence $\ket{\PsiDet}, \Ham\ket{\PsiDet}, \Ham^2\ket{\PsiDet}, \hdots$, then the procedure terminates at $k=w$ with $\ket{\beta_w} = 0$.
  Although it can give the exact result in principal, the classical Gram-Schmidt procedure is cumbersome in that it requires $\sSum{k=1}{w-1} k = w(w-1)/2$ orthogonalization operations.
  The dimension $w$ of the bright space might be much smaller than the dimension of the total Hilbert space, but $w$ can nevertheless be large.
  In this case we are still faced with tremendous efforts.

  We say that the state $\ket{\psi}$ has distance $s$ from the detection state, if $\mel*{\psi}{\Ham^s}{\PsiDet} \ne 0$ but $\mel*{\psi}{\Ham^{s'}}{\PsiDet} = 0$ for all $0\le s' < s$.
  With this notion of distance, the Hamiltonian and the detection state induce a ``shell structure'' in the Hilbert state.
  Fixing any basis $\{ \ket{r} \}$ of the Hilbert space, the $s$-th shell consists of all basis states that have distance $s$ from $\ket{\PsiDet}$.
  These are the states on which the state $\Ham^s \ket{\PsiDet}$ is supported and which do not belong to a previous shell.
  This picture is especially intuitive in quantum walks with the graph node basis.
  Here, our notion of distance coincides with the graph distance and is particularly useful.
  The topology of the graph translates naturally into the shells' connections.
  However, our notion is not restricted to quantum walks on graphs.

  From this ``shell'' perspective, it is clear that the first Gram-Schmidt vector that contributes to the total detection probability must be the $\xi$-th, where $\xi$ is the distance between the initial and the detection state.
  The first non-trivial lower bound from the Gram-Schmidt procedure thus requires $\xi(\xi-1)/2 = \sLandau{\xi}$ orthogonalization operations.
  
  We therefore propose a simpler procedure that only requires $\sLandau{\xi}$ operations.
  We try to iteratively construct a bright state that is concentrated in the $\xi$-shell alone.
  We base our technique on the idea that the $s$-th shell is usually only connected to the $(s+1)$-th and the $(s-1)$-th shell.
  Therefore, it is to be expected that the overlap between each shell with its two precedessors dominates.
  This overlap is erased by orthogonalization in each step and gives $\ket*{\tilde{\beta}_s}$.
  $\Ham\ket*{\tilde{\beta}_s}$ is used for the next step (as opposed to $\Ham^{s+1}\ket{\PsiDet}$), so that $\ket*{\tilde{\beta}_{s+1}}$ benefits from the previous steps' erasures.
  In terms of an equation, we use:
  \begin{equation}
    \ket*{\tilde{\beta}_k} 
    := 
    \tilde{N}_k
    \Prod{k'=k-2}{k-1}
    ( \Id - \dyad*{\tilde{\beta}_{k'}} )
    \Ham \ket*{\tilde{\beta}_{k-1}}
    ,
  \label{eq:}
  \end{equation}
  as well as $\ket*{\tilde{\beta}_1} = \Ham\ket{\PsiDet}$, and $\ket*{\tilde{\beta}_0} = \ket{\PsiDet}$.
  All these states can be trivially obtained from iteration.
  Each step requires the same amount of effort as the last.
  All of them are bright, as they are superpositions of the bright states $\{ \Ham^k\ket{\PsiDet} \}$.
  However, only selected pairs are orthogonal to each other (namely the subsequent ones).
  Note that although we alluded to a special choice of basis in the beginning, these shell states do not depend any basis.
  They can be computed for any system.

  For an initial state $\ket{\PsiIn}$ at distance $\xi$ from $\ket{\PsiDet}$ we obtain the following lower bound:
  \begin{equation}
    \TDP(\PsiIn)
    \ge
    \abs*{\ip*{\tilde{\beta}_\xi}{\PsiIn}}^2
    ,
  \label{eq:}
  \end{equation}
  which is obtained from only $1 + 2\xi$ orthogonalizations.

  For the hypercube this procedure yields
  \begin{equation}
    \ket*{\tilde{\beta}_k} = \binom{B}{k}^{-\frac{1}{2}} \Sum{r_k}{} \ket{r_k}
    ,
  \label{eq:}
  \end{equation}
  and thus the lower bound coincides with the exact result $\TDP(r_\xi) \ge 1/\binom{B}{\xi}$.
  Similar for the ring, where $\ket*{\tilde{\beta}_k} = ( \ket{d+k} + \ket{d-k} )/\sqrt{2}$ and $\TDP(d\pm\xi) \ge 1/2$.

  For bipartite graphs, even and odd $\ket*{\tilde{\beta}_k}$ decouple and it suffices to only look at one of these subsets.
  This lower bound saturates, i.e. gives the exact result, whenever the overlaps $\ip*{\tilde{\beta}_k}{\tilde{\beta}_{k+j}}$ vanish anyhow for $\abs{j} >2$.
  (Then, we actually did not omit any orthogonalization operation.)
  This is the case for the hypercube, the line graph when the detection state is localized on the end of the line, and for the tree, when the detector resides on the root.
  However, this is not generally the case.
  In particular, the such obtained lower bound can not be obtained when the detection state is a superposition supported on non-neighboring sites.

\section{Different lower bounds for the hypercube}
  \begin{figure}
  \centering
  \includegraphics[width=0.99\columnwidth]{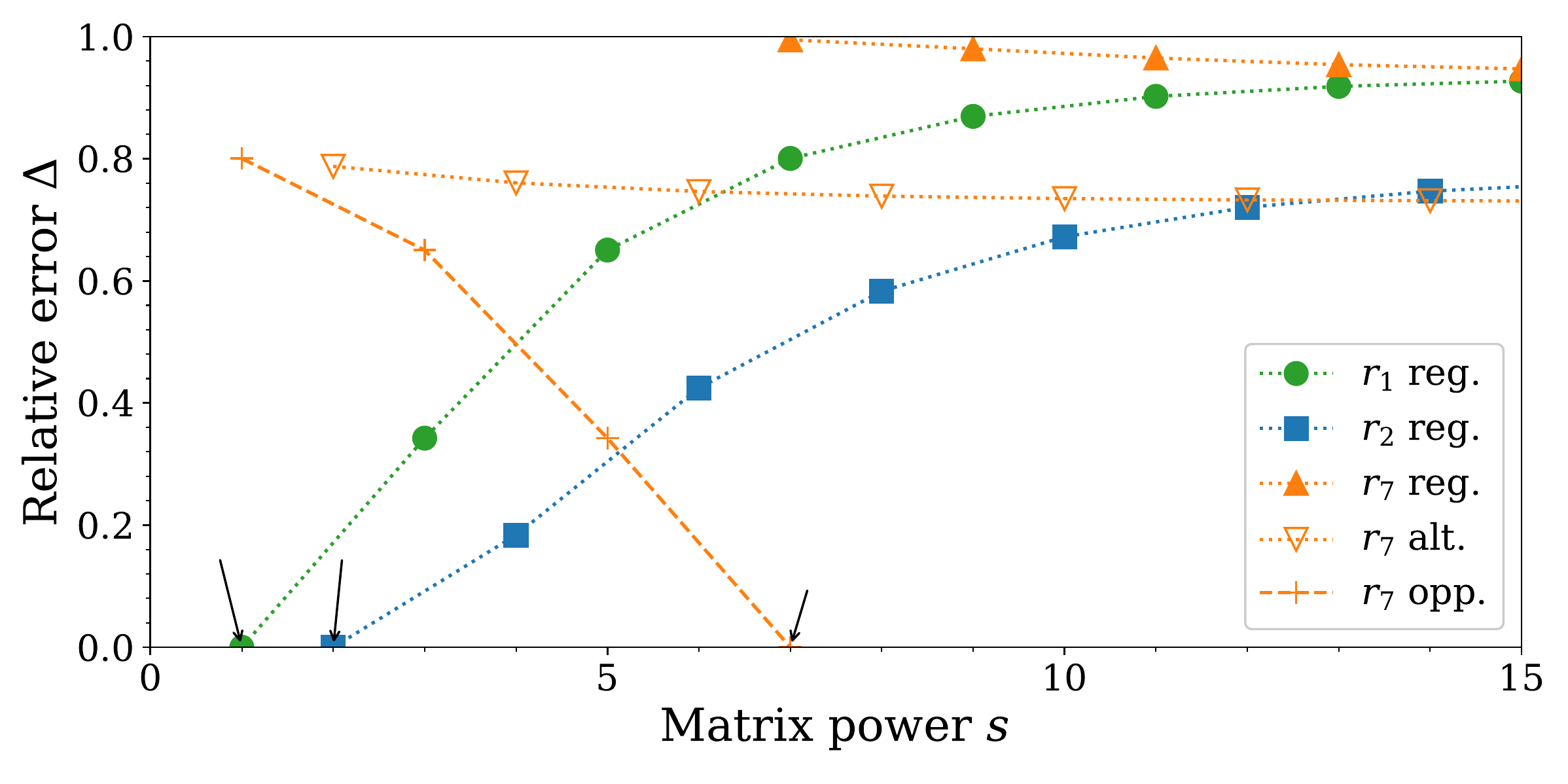}
  \caption{
    The relative error versus $s$ in a hypercube with $B=8$.
    We vary $s$ to find the tightest bounds (arrows).
    The initial sites $\ket{r_\xi}$ are $\xi = 1,2,7$  nodes away from the detection site.
    Exact result and symmetry bound coincide $\TDP(r_\xi) = 1/\nu = 1/\binom{B}{\xi}$.
    Every second $s$ is omitted, because the hypercube is bipartite.
    For small distances ($\xi=1,2$), $s=\xi$ gives coincident bounds for the regular strategy.
    For large distances ($\xi=7$), the bounds are loose.
    The alternative strategy gives a slightly better bound.
    The opposite strategy gives again a lower bound that coincides with the exact result.
  }
  \label{fig:Hypercube}
  \end{figure}
  \begin{table}
    \begin{tabular}{l|c|c|c}
           & $\ket{\beta_1}$ & $\ket{\beta_2}$ & Best $\Delta$ \\
      \hline
      reg. & $\ket{\PsiDet}$ & $(\Id - \dyad{\PsiDet}) \Ham^s \ket{\PsiDet}$ & $0.95$ \\
      alt. & $\Ham^\xi\ket{\PsiDet}$ & $(\Id - \dyad{\beta_1})\Ham^{s+\xi}\ket{\PsiDet}$ & $0.77$ \\
      opp. & $\ket{r_B}$ & $(\Id - \dyad{r_B})\Ham^{B-s}\ket{r_B}$ & $0.0$ \\
      opt. & $\Ham^{s_1}\ket{\PsiDet}$ & $(\Id - \dyad{\beta_1})\Ham^{s_2}\ket{\PsiDet}$ & $0.66$

    \end{tabular}
    \caption{Different strategies to get a lower bound \label{tab:Hypercube}}
  \end{table}
  Here, we compare the different lower bounds that have been mentioned in the main text in a hypercube with $B=8$ and localized initial and detection states.
  These are obtained from the regular (reg.), alternative (alt.), opposite (opp.), and the optimization (opt.) strategies.
  They all are based on a different choice of initial bright states $\ket{\beta_1}$ and $\ket{\beta_2}$ in Eq.~(2) of the main text.
  These choices are summarized in Table~\ref{tab:Hypercube} (up to normalization).
  Repeating the main text's procedure, one obtaines a lower bound that depends on the free parameters $s$ or $s_1$ and $s_2$.
  These can then be tunede to find the largest (thus best) lower bound.
  We compare these methods with respect to the relative error of the bounds to the exact total detection probability.
  \begin{equation}
    \Delta 
    :=
    \frac{\TDP^\text{upp} - \TDP^\text{low}}{\TDP}
    .
  \label{eq:}
  \end{equation}
  Keep in mind that the upper, symmetry bound coincides with the exact result $\TDP^\text{upp}(r_\xi) = \TDP(r_\xi) = 1/\binom{B}{\xi}$, which is a merit of the system's high symmetry.
  The results are depicted in Fig.~\ref{fig:Hypercube}.
  For small $\xi$, the regular strategy gives vanishing relative error.
  This means that the lower and upper bound coincide and thus determine the exact result.
  Large $\xi$ yield a loose bound.
  The best $\Delta$ values of each strategy for the initial state $\ket{r_7}$ are summarized in the last column of Table~\ref{tab:Hypercube}.
  The regular strategy can be slightly improved upon by the alternative or optimization strategy.
  (We found the optimal values to be $s_1 = 17$ and $s_2 = 19$.
    Larger parameters do give lower $\Delta$s but only in the third digit.)
  The opposite strategy on the other hand gives again a vanishing relative error and pins down the exact total detection probability.
  The same holds for the aforementioned modified Gram-Schmidt procedure.

\bibliographystyle{apsrev}

\end{document}